\documentstyle[12pt]{article}
\advance \textheight by \topskip
\begin{document}

\begin{titlepage}
\hbox to \hsize{\hfil {\bf hep-ph/9712262 }}
\hbox to \hsize{\hfil }
\hbox to \hsize{\hfil }
\vfill
\large
\begin{center}
{\bf Spin asymmetries at RHIC and nonperturbative aspects\\
of hadron dynamics}
\end{center}
\vskip 1cm
\normalsize
\begin{center}
{\bf S. M. Troshin and N. E. Tyurin}\\[1ex]
{\small\it Institute for High Energy Physics,\\
  Protvino, Moscow Region, 142284 Russia}
\end{center}
\vskip 2.cm
\begin{abstract}
Some nonperturbative aspects of spin studies at RHIC are discussed
and the predictions for single- and two-spin asymmetries are
given.  Among them are those which
emphasize the role of angular orbital momentum in the spin
structure of the constituent quarks.
\end{abstract}
\vfill
\end{titlepage}

\section*{Introduction}
The  spin studies program at RHIC
has several well defined goals and among them:
\begin{itemize}
\item
study of the nucleon spin structure, i.e., how the proton's spin
state can be obtained from a superposition of Fock states with different
numbers of constituents with nonzero spin;
\item
study how the dynamics of constituent interactions depend on  spin;
\item
understanding the chiral symmetry breaking and helicity non-conservation on 
quark and hadron levels;
\item
study of the overall nucleon structure and long range dynamics.
\end{itemize}

 These issues are closely interrelated at the hadron level and
 the results of the experimental measurements 
 are to be interpreted in terms of hadron  spin
structure convoluted with the constituent interaction dynamics.

In this paper we discuss the respective physics potential
of RHIC machine.

\section{Single-spin asymmetries in inclusive processes}
 Studies of the
 spin effects in
inclusive processes probe the  spin dependence of
 the incoherent hadronic interaction dynamics.
The cross--sections of the hard production processes
are described in the
perturbative QCD as a convolution integral of parton
cross-sections with the light--cone parton densities.

These studies  would yield
information on the contribution of the spin and orbital
angular momenta of  quarks and gluons
into the  hadron helicity:
\begin{equation}
1/2=1/2\Delta\Sigma+L_q+\Gamma+L_g
\end{equation}
In the above sum  all  terms have clear physical interpretation,
however  besides the first one, they are gauge
 and frame dependent. Transparent discussions of the theoretical aspects
of this  sum rule  are given in
 \cite{jin}.

 The primary goal of the single-spin measurements
 with hadronic final states
 would be a study where the onset of perturbative QCD
regime occurs. In the spin measurements in inclusive process
$A+B\to C+X$ with polarized hadron $A$
this is based on the assumption of the higher--twist origin of the one--spin
transverse asymmetries \cite{kane,efremt}.
The contribution of higher twists should be small at high energies
where the interactions at small distances $l\sim 1/Q$ can be studied.
There are some indications that such contributions are small even at
not too high energies and $Q^2$ values.
In particular, it follows
from the recent data on the spin structure function
$g_2(x)$ obtained at SLAC. If it is the case, the observed
significant one-spin asymmetries in hadronic processes are to be
associated with 
manifestation of  nonperturbative dynamics.
However, the available
energies are not high enough to make the unambiguous conclusion.
Therefore, the measurements of one-spin asymmetries at RHIC
energies are crucial. For the production of hadrons with high
$p_\perp$
        the simple
factorization is valid for the transversely polarized hadrons as well
as for the longitudinally polarized ones \cite{coll}, i.e.
\begin{equation}
A_N d\sigma\sim \sum_{ab\rightarrow cd}\int d\xi_Ad\xi_B\frac{dz}{z}\delta
f_{a/A}(\xi_A) f_{b/B}(\xi_B)\hat a_N d\hat \sigma
_{ab\rightarrow cd} D_{C/c}(z) \end{equation} and we are  expecting
vanishing one--spin transverse asymmetries \[ A_N=0.  \]
since at the leading twist level $\hat a_N= 0$.

It can also be demonstrated expressing the asymmetry $A_N$ through
the helicity
amplitudes. Indeed,  asymmetry $A_N$ results from the interference
between helicity amplitudes which differ by one unit of helicity
\begin{equation}
A_N=2\frac{\sum_{X,\lambda _X, \lambda _2}\int d\Gamma _X
\mbox{Im}[F_{\lambda _X;+,\lambda _2}F^*_{\lambda_X;-,\lambda _2}]}
{\sum_{X,\lambda _X;\lambda _1 \lambda _2}\int d\Gamma _X
|F_{\lambda_X;\lambda _1 ,\lambda _2}|^2},
\end{equation}
where $F_{\lambda _i}$ are the helicity amplitudes of exclusive
processes. If the helicity conservation in QCD for exclusive
processes is valid at hadron level \cite{brlp}
\begin{equation}
\lambda_1+\lambda_2=\lambda_X
\end{equation}
 then it follows that $A_N=0$
 in the phase  when chiral symmetry is not broken,
 i.e. in the limit of high $p_{\perp}$'s. Note, that 
 $\lambda_X=\sum_{i=1}^{n_X}\lambda_i$ and $n_X$ is the number of particles
 in the exclusive final state $X$. Essential point here is the assumption
 that at the short distances vacuum is a perturbative one.
Thus, the study of $p_{\perp}$--dependence of one--spin
asymmetries might be used as a way to reveal the transition from the
non--pertur\-ba\-tive phase ($A_N\neq 0$) to the perturbative
phase ($A_N=0$) of QCD. The very existence of such transition can
 not be taken
for granted since the vacuum, even at
short distances, could be filled with the fluctuations of gluon or quark
fields \cite{prep}. 

The measurements of one--spin transverse
asymmetries in this case will be important probe of the
 chiral structure of the effective QCD Lagrangian.

It should be mentioned that available experimental data 
are at some variance with PQCD predictions: the data 
do not at least show up
tendency to converge to the vanishing single-spin asymmetries
in inclusive and elastic hadron productions.
Indeed, in the soft hadronic interactions  significant single-spin 
effects could be expected since the helicity conservation rule
does not work in the interactions at large distances
 because the chiral $SU(3)_L\times SU(3)_R$ symmetry of 
 the QCD Lagrangian is spontaneously broken.
However, the
asymmetries at low transverse momentum are small.
On the other hand, these asymmetries increase at
higher transverse momentum; but 
this is just where we should expect
$A_N=0$. 

Several mechanisms have been proposed for the explanation
of the observed single-spin effects.
Possible  sources of the observed one-spin asymmetries could be:
higher
twist effects \cite{efremt}, correlation of $k_\perp$ and spin in
structure \cite{sivs} and fragmentation \cite{cols,artru} functions,
rotation of valence quarks inside a hadron \cite{boros} and the
coherent rotation of the quark matter inside the constituent
quarks \cite{asy}.

The main points of the mechanism proposed in \cite{asy} are:
\begin{itemize} \item asymmetry in the pion production
reflects the internal spin structure
of the constituent quarks, i.e. it arises due to the orbital angular
momentum of current quarks inside the constituent quark; \item sign
of asymmetry and its value are proportional to polarization of the
 constituent quark inside the polarized nucleon.
\end{itemize}

The model predicts significant one-spin asymmetries at high
$p_{\perp}$ values.
The significant asymmetries appear to show up beyond
$p_{\perp}>\Lambda _\chi(\simeq 1-2$ GeV/c), i.e.
the scale where internal structure of a constituent
quark could be probed (Fig. 1).

The observed $p_{\perp}$-behavior of
asymmetries in inclusive processes  confirms these predictions
\cite{brav} and indicates that the transverse size $\Delta b$ of the region
where the asymmetries originate from is significantly less than the
hadron size $\Delta b_H$ and is of the order of the magnitude of
the size of constituent quark, i.e.
\[
\Delta b\sim 1/m_Q.
\]

The relevant processes for the study of above problems are the
following:
\begin{equation}
p_{\uparrow}+{p}\rightarrow \pi^{} ,\gamma ,\mbox{jet}+X
\end{equation}

As an example we consider  charged pion production.
 The model \cite{asy}
predicts
\begin{equation} A_N(s,x,p_{\perp})=\frac{ \sin[{\cal{P}}_{\tilde Q}(x)
\langle L_{\{\bar q q\}}\rangle] {W_+^h(s,\xi)}}{ {
[W_+^s (s,\xi)+W_+^h(s,\xi)]}},
 \end{equation}
where the  functions $
W_+^{s,h}$ are determined by the interactions at large and small
distances.
 
The asymmetries have energy independent values about 
30\% at $p_\perp>\Lambda_\chi
(\simeq 1-2$ GeV/c) and $x\simeq 0.5$:
\[
A_N^{\pi^\pm}\simeq\pm 30 \% .
\]
Note, that the corresponding asymmetry in the neutral pion
production has value about $5-6\%$ in the same kinematical
region.

As it has been noted higher twists  at RHIC energies and
high $p_\perp$'s would not provide significant contribution to asymmetry
$A_N$. Phenomenology of such contributions has not been developed yet
and predictions could vary from a few  to tens percents. However,
higher twists provide a decreasing single-spin asymmetry at 
high $p_\perp$
\[
A_N^{\pi^\pm}\sim M/p_\perp.
\]
 The same is true for the predictions
of the mechanisms discussed in \cite{sivs}.

Experimental error in asymmetry $A_N$ is given by the events
number $N$ and the beam polarization $P_1$:
\begin{equation}
\delta A =\frac{1}{P_1}\frac{1}{\sqrt{N}}.
\end{equation}
For the values of RHIC luminosity $L=2\cdot 10^{32}$
 $cm^{-2}s^{-1}$ and beam polarization $P_1=70\%$
the experimental measurements of single-spin asymmetry in pion
production with accuracy of a few percents are feasiable up
to $p_\perp$ about 30 GeV/c \cite{tan}.
It would be important to study the processes of the charged
pion production since the available data at $p_L=200$ GeV/c
indicated that the observed symmetries could be of order
of tens percents.

The $p_L=200$ GeV/c is the maximal energy
the asymmetries in these processes has been measured so far.
The study of this reaction at RHIC energies could provide  a clear
test of perturbative QCD regime and nonperturbative  models.
In general, these studies are important for understanding
of the QCD vaccuum and transitions between the perturbative
and nonperturbative phases.

In the $\pi^0$ production the asymmetries are expected to be smaller
due to the isospin substraction between  $\pi^\pm$ cross sections.

\section{Transverse spin densities and two--spin correlations
in inclusive processes}

The transverse quark spin density measures the difference of
 the quark momentum
distributions in a transversely polarized nucleon when a quark is
polarized parallel or antiparallel to the nucleon. This quantity
$\delta q(x)$, unlike the longitudinal quark spin
density $\Delta q(x)$,
cannot be measured in deep inelastic scattering due to its different
properties under chiral transformations.
Extensive studies of the theoretical
aspects of transverse spin
structure of the nucleon were made in  \cite{rlsp} and \cite{arme,jfji,trsv}.
Quark transversity is a new observable for understanding
the hadron
wave function in terms of bare quarks. Gluons give no contribution
to the transverse spin of the proton. It is promising to explore this
new spin observable and compare it with the longitudinal spin densities.

Recently an upper
 bound for $\delta q(x)$ 
\cite{sof}:
\begin{equation}
|\delta q(x)|\leq q_+(x)
\end{equation}
and the bound for $\delta q(x)$ limiting its behavior at $x\to 0$ \cite{tro}:
\begin{equation}
\delta q(x)\leq\log{x}/x
\end{equation}
have been obtained.

Low-$x$ behavior of the spin structure functions
$g_1(x)$ and $h_1(x)$ has been considered \cite{dif} in the unitarized chiral
quark model which
combines ideas on the constituent quark structure of hadrons with
a geometrical scattering picture and unitarity.  A nondiffractive
singular low-$x$  dependence of $g^p_1(x)$ and $g_1^n(x)$ 
was obtained and a diffractive type
smooth behaviour of $h_1(x)$  is  predicted at small $x$. 
 The quark densities $q(x)$, $\Delta q(x)$ and $\delta q(x)$ 
 at small $x \simeq Q^2/s$ have the following forms \cite{dif}:
\begin{equation}
q(x)\sim \frac{1}{x}\ln^2(1/x)\label{df1},
\end{equation}
\begin{equation}
\Delta q(x)\sim \frac{1}{\sqrt{x}}\ln(1/x)\label{dg1}
\end{equation}
and
\begin{equation}
\delta q(x)\sim x^{\frac{\alpha-1}{n+1}}\ln(1/x).\label{dh1}
\end{equation}
It also follows that the unpolarized structure function $F_1$
and transversity $h_1$
will be the universal for the proton and neutron, i.e.:
\begin{equation}
F^p_1(x)=F_1^n(x)\sim
\frac{1}{x} \ln^2(1/x).\label{f1m}
\end{equation}
and
\begin{equation}
h^p_1(x)=h_1^n(x)\sim
 x^{\frac{\alpha-1}{n+1}}\ln(1/x).\label{h1m}
\end{equation}
It is also seen that $h_1(x)$ has a smooth behaviour at $x\to 0$, i.e.
$h_1(x)\to 0$ in this limit ($\alpha>1$).

So far the Drell--Yan process with transversely polarized protons in the
initial state
\[
p_{\uparrow}+p_{\uparrow}\rightarrow \mu ^+ + \mu ^- + X
\]
is
most suitable for the determination of the quark transversity.
The corresponding two--spin asymmetry is directly
 related to the quark
transversity distributions $\delta q(x)$.
The explicit low-$x$ forms
Eqs. (\ref{dg1}) and (\ref{dh1}) allow one to analyze 
the asymmetries $A_{TT}$ and $A_{LL}$ in the central
region of the low-mass Drell-Yan production.
They appear to be small  at $x_F\simeq 0$ 
\begin{equation}
A_{LL}^{l\bar l}\simeq 0\quad
\mbox{and}\quad
A_{TT}^{l\bar l}\simeq 0
\end{equation}
when invariant mass of the lepton pair
$M^2_{l\bar l}\ll s$
At the same time the ratio of the asymmetries
$A_{TT}^{l\bar l}$ and $A_{LL}^{l\bar l}$ is also small in this
kinematical region:
\begin{equation}
A_{TT}^{l\bar l}/A_{LL}^{l\bar l}\simeq 0.
\end{equation}
This result agrees with the predictions
made in \cite{sait}. Despite the small predicted asymmetries
the experimental measurements
in this kinematical region could be important.

The measurements
of the two--spin longitudinal
asymmetries will probe the gluon contribution $\Delta g(x)$
 to the helicity of
the nucleon. The relevant processes for that purpose are the direct
$\gamma $, jets, $\chi _2$ and pion--production at high
$p_\perp$'s in the collisions of longitudinally
polarized protons.
Estimation of the experimental sensitivity $\delta A_{LL/TT}$
shows that the measurements
with accuracy of a few percents are feasiable at RHIC up
to $p_\perp$ about 30 GeV/c.

\section{Strangeness in the hadrons}
It is  evident from deep--inelastic
scattering data that
strange quarks alongside with gluons could play essential role in
the spin structure of nucleon.
 DIS data show that strange quarks are
 negatively polarized in polarized
nucleon, $\Delta s\simeq -0.1$.
Elastic $\nu p$-scattering data as well provide the value
 $\Delta s=-0.15\pm 0.08$ \cite{nu}.
The presence and polarization of strange quarks inside a hadron should
give an experimental signal in hadronic reactions as well.

We address now the asymmetry in
 the  production of
 $\varphi$-meson consisting of strange quarks.
It is worth to stress that in addition to $u$
 and $d$ quarks the constituent quark ($U$, for example) contains pairs
of strange quarks  and the ratio of scalar
density matrix elements \begin{equation} y=
{\langle U| \bar ss|U\rangle}
/ {\langle U|\bar u u+\bar d d+\bar s s|U\rangle}
 \label{str} \end{equation} is estimated  as $y=0.1 - 0.5$.

 It was argued \cite{str} that the single spin  asymmetry $A_N$ in the
process $pp\to \varphi X$
 is due to orbital momenta of
 strange quarks in the internal structure of constituent quarks.
The  estimation for the asymmetry $A_N$ in $\varphi$-meson
production at $p_\perp>\Lambda_\chi(\simeq 1-2$ GeV/c) is:
\begin{equation}
A_N(\varphi)\propto \langle{\cal{P}}_{ Q}\rangle \langle L_{\{\bar q
q\}}\rangle y\simeq 0.01-0.05.
  \label{an} \end{equation}
Thus,  a quite noticeable  one-spin asymmetry at high $p_{\perp}$
values in inclusive $\varphi$-meson production can be expected.
The above estimate also shows that it
 is reasonable to make experimental measurements of the cross-section and
asymmetry in inclusive $\varphi$-meson production to study strange content of
constituent quark as a possible source of OZI-rule evasion.

As it now is  known, only part (less
than one third in fact) of the proton spin is due to quark spins
\cite{ellis,altar}.  These results one can  interpret in the
effective QCD approach ascribing a substantial part of hadron spin
to an orbital angular momentum of quark matter.
This orbital angular momentum might be revealed when
asymmetries in hadron production are measured. 
The main role  belongs to the orbital angular
momentum of $\bar q q$--pairs inside the constituent quark while
constituent quarks themselves have very slow (if at all) orbital
motion and the hadron wave function may be  approximated 
by $S$-state of the
constituent quark system.  The observed $p_{\perp}$--dependence of
$\Lambda$--hyperon polarization  in inclusive processes seems
to confirm such conclusions, since
it  appears to show up beyond
 $p_\perp>\Lambda_\chi(\simeq 1-2$ GeV/c)
 i.e.  the scale where internal structure of  constituent
quark can be probed (Fig. 2).
    The main outcome of the considered model:
polarization of
$\Lambda$ -- hyperons arises as a result of the  internal structure of the
constituent quark and its multiple scattering in the some effective  field. It is
proportional to the orbital angular momentum of strange quarks  which is 
initially resided inside the
constituent quark.
It is predicted in this model that the double spin correlation
parameters should
have a similar $p_\perp$-dependence:
\begin{equation}
D_{TT}\sim D_{LL}\sim 0\label{dl1}
\end{equation}
at
 $p_\perp<\Lambda_\chi(\simeq 1-2$ GeV/c)
and
\begin{equation}
D_{TT}\sim D_{LL}\sim \mbox{const.}\label{dl2}
\end{equation}
at
 $p_\perp>\Lambda_\chi $
in the polarized beam fragmentation region in the processes
\[
p_\uparrow+p\to \Lambda_\uparrow + X
\]
and
\[
p_\rightarrow+p\to \Lambda_\rightarrow + X.
\]
Eqs. (\ref{dl1}), (\ref{dl2}) reflect the fact that the polarized strange quark
is located
inside the constituent quark of a smaller than a hadron size.

It is the generic feature of the above refenced  model: spin effects in inclusive
processes are related to
the internal structure of constituent quark. This fact could explain
similarity in the observed behaviour of different spin observables
in inclusive
processes, i.e. its rise with $p_\perp$ at small and medium transverse
momenta and then flattening at higher values of $p_\perp$.

It would be interesting to check these predictions at RHIC
energies as well as to measure for the first time triple spin
correlation parameters in the processes of hyperon production
with two polarized proton beams. It would help to understand the
mechanism of hyperon polarization.

\section*{Conclusion}

We have considered above the inclusive processes. 
Spin measurements at RHIC with single and both polarized proton beams
would probe the fundamental couplings of the underlying Lagrangian
and investigate the spin structure of the nucleon.
 A variety  of one-- and two--spin asymmetries could be
measured. As it has often happened in the past,
these spin measurements might bring unexpected new results;
this would certainly stimulate the development of new
theoretical ideas.

Among the spin studies the most exiting ones are those related
to the nonperturbative effects. Asymmetries in fixed angle elastic
scattering are predicted to have significant values  by several
models based on the nonperturbative dynamics \cite{ttel,mqm,diq}.
One should also say that despite the obvious experimental difficulties in conducting elastic
scattering experiments the data would be extremely desirable for the
comprehensive spin physics program

\vspace{0.5cm}
We are pleased to thank Mike Tannenbaum for his helpful comments and
reading the manuscript.

\newpage \normalsize \begin{center} \bf Figure captions \end{center}
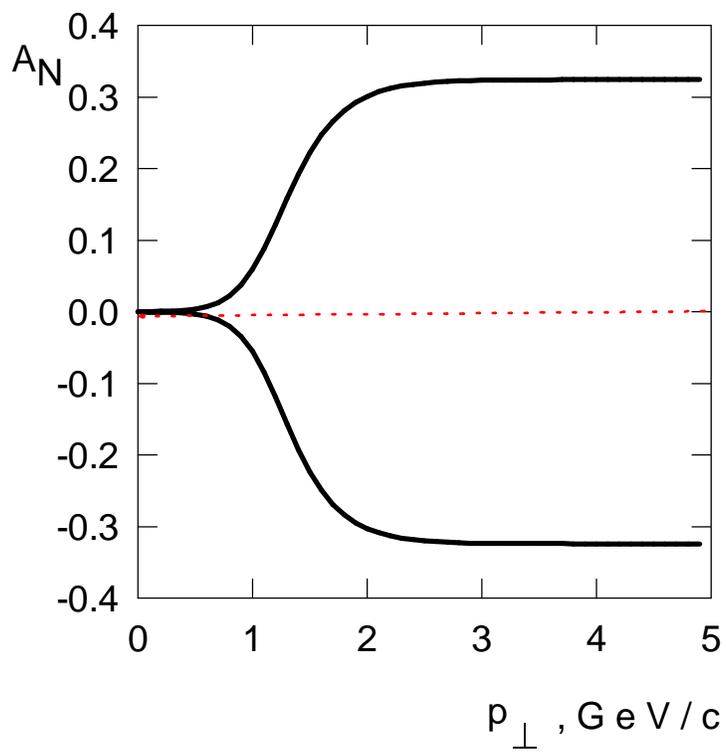
\rm \bf Fig. 1. \rm Asymmetry $A_N$ in the process
$p_{\uparrow}+p\rightarrow \pi^++X$ (positive values) and in the
process $p_{\uparrow}+p\rightarrow \pi^-+X$ (negative values) at
$\sqrt{s}=500$ GeV.\\[2ex]
\bf Fig. 2 \rm The $p_\perp$--dependence of $\Lambda$--hyperon polarization
in the process $pp\rightarrow\Lambda X$ at $p_L=400$ GeV/c.
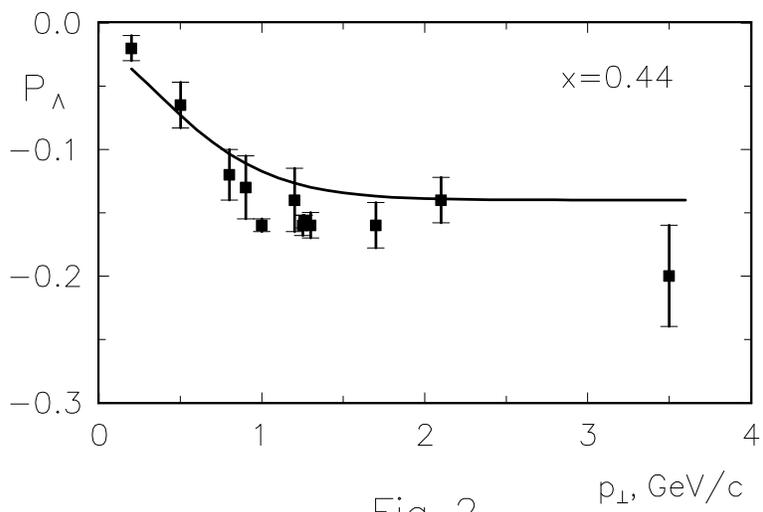
\end{document}